# Thermodynamic assessment of machine learning models for solid-state synthesis prediction

Jane Schlesinger[1], Simon Hjaltason[1], Nathan J. Szymanski[1], Christopher J. Bartel[1*]

[1] University of Minnesota, Chemical Engineering and Materials Science, Minneapolis, MN 55455

[*] correspondence to cbartel@umn.edu

## Abstract

Machine learning models have recently emerged to predict whether hypothetical solid-state materials can be synthesized. These models aim to circumvent direct first-principles modeling of solid-state phase transformations, instead learning from large databases of successfully synthesized materials. Here, we assess the alignment of several recently introduced synthesis prediction models with material and reaction thermodynamics, quantified by the energy with respect to the convex hull and a metric accounting for thermodynamic selectivity of enumerated synthesis reactions. A dataset of successful synthesis recipes was used to determine the likely bounds on both quantities beyond which materials can be deemed unlikely to be synthesized. With these bounds as context, thermodynamic quantities were computed using the CHGNet foundation potential for thousands of new hypothetical materials generated using the Chemeleon generative model. Four recently published machine learning models for synthesizability prediction were applied to this same dataset, and the resultant predictions were considered against computed thermodynamics. We find these models generally overpredict the likelihood of synthesis, but some model scores do trend with thermodynamic heuristics, assigning lower scores to materials that are less stable or do not have an available synthesis recipe that is calculated to be thermodynamically selective. In total, this work identifies existing gaps in machine learning models for materials synthesis and introduces a new approach to assess their quality in the absence of extensive negative examples (failed syntheses).

## Introduction

Inorganic crystalline solids underpin numerous technologies, motivating the discovery of new materials with optimized properties to support further innovation. Density functional theory (DFT) calculations are now regularly applied to compute materials properties in high throughput, allowing candidates to be screened prior to synthesis.[1–4] Meanwhile, generative artificial intelligence (AI) models are being used to propose thousands of novel materials with desired properties, further expanding the pool of hypothetical materials.[5] To support this effort, researchers often leverage calculable thermodynamic heuristics as a filter to identify viable and realistic synthesis candidates. Most commonly used is thermodynamic stability, which can be quantified by the energy of a material above the convex hull ($E_{hull}$) defined by known ground states. $E_{hull}$ is a relative measure obtained by constructing a phase diagram for the chemical space of interest. The energies used in this construction can be drawn from existing databases[6–9] or computed using density functional theory (DFT) or interatomic potentials for the materials of interest.[10–14] If a given material at a particular composition cannot lower its energy by phase separation or phase transition, it is stable ($E_{hull}$ = 0) and therefore likely to be synthesizable. Thermodynamically unstable materials can often still be synthesized, but those near the hull (low $E_{hull}$) are generally considered more accessible.[15] Sun *et al.* found that about 50% of all experimentally reported materials in the Inorganic Crystal Structure Database (ICSD)[16] are stable and 90% of those that are unstable (above the hull) exhibit $E_{hull} \leq$ 67 meV/atom as computed with DFT at 0 K.[17]

Thermodynamic stability is effective as a preliminary filter of candidate materials, but turning these feasible candidates into experimentally made materials remains a major challenge. As such, the number of hypothetical materials (e.g., those appearing in open DFT databases[4,18]) far outpaces those that have been experimentally realized (e.g., those appearing in the ICSD). The core bottleneck is our inability to predict whether and how a material can be synthesized.[19] Thermodynamically stable materials should form at equilibrium if the conditions used for synthesis are not too dissimilar from those accounted for computationally. However, an actual solid-state reaction may not reach equilibrium, or it is possible that synthesis conditions (e.g., temperature) have a significant influence on the resulting thermodynamics. The product that does form is determined by the reaction pathway, which is itself dependent on the concerted motion of many atoms through defect formation, ion transport, nucleation, and crystal growth, among other processes. The complex mechanism by which crystalline materials transform precludes direct atomistic modeling of the synthesis process, and solid-state reaction outcomes are therefore challenging to directly predict. As a result, costly trial-and-error experimentation is often an unavoidable, and sometimes fruitless, step in the discovery of novel materials.

The successful synthesis of a new material depends on the chosen synthesis recipe, defined by the precursors and reaction conditions. There have been several recent efforts to apply first-principles thermodynamic calculations to anticipate how the choice of synthesis recipe influences the likely phase evolution sequence, mostly in the context of solid-state synthesis.[20–24] Towards the goal of identifying "optimal" solid-state synthesis recipes, McDermott *et al.* formalized an approach to enumerate and quantify the plausibility of many pathways to a given target. This approach enables the identification of precursors that minimize a user-specified cost function (e.g., related to the anticipated reaction energies) to form the desired target.[25] Building on this work, the authors posited that optimal precursors have both a strong driving force (i.e., a large and negative Gibbs free energy of reaction, $\Delta G$) to form the target material and minimal driving force for "competing" impurity-forming reactions.[26] In this context, primary competition ($C_1$) is defined as the difference between $\Delta G$ to form the target and $\Delta G$ to form the most favorable competing product from a given precursor set. Secondary competition ($C_2$) measures the tendency of the target material (once formed) to react with unconsumed precursors and produce unwanted impurities. The authors proposed an aggregate metric ($\Gamma$) to account for both $C_1$ and $C_2$:

$$\Gamma = w_0 \Delta G_{\mathrm{rxn}} + w_1 C_1 + w_2 C_2 \tag{1}$$

$\Delta G_{\mathrm{rxn}}$ measures the reaction energy associated with target formation from the precursors, and the weights ($w_0$, $w_1$, $w_2$) can be chosen to weight the influence of reaction energy, primary competition, and secondary competition as desired. The choice of precursors that minimizes $\Gamma$ for a given target is then deemed the optimally selective recipe for that target from a thermodynamic perspective. Unlike $E_{\mathrm{hull}}$, which is defined for a material, $\Gamma$ is defined for a chosen synthesis recipe, so the two quantities should be complementary in the discovery of new materials.

Machine learning (ML) methods are emerging as an alternative approach to predict the synthesizability of hypothetical materials without the need for first-principles thermodynamic calculations. ML models can capture complex, often elusive, physical relationships making them a promising tool for the multifaceted challenge of predictive synthesis. Existing applications of ML for synthesis prediction have focused on predicting whether a hypothetical material can be synthesized by any approach, irrespective of available synthesis recipes. In this way, models are tasked with labeling materials as either synthesizable or unsynthesizable. The success of ML is generally contingent on sufficient data availability. While the literature provides extensive data on successfully synthesized materials and the synthesis recipes used to make them,[16,27] it is more difficult to obtain reliable "negative" data.[28] The absence of a material from these databases does not imply it is unsynthesizable; rather, it might not have been attempted or reported yet. Without clear negative examples, conventional classification models cannot learn the distinction between synthesizable and unsynthesizable materials.

Jang *et al.* proposed a positive-unlabeled (PU) learning approach to address the limited availability of negative synthesis data. The authors employed PU learning to assign labels to unreported materials with a transductive bagging approach, aggregating the labels assigned by models trained on different subsets of the data to ultimately label all unlabeled materials.[29] Experimentally reported materials are treated as positive examples, and the method for labeling assumes that materials differing substantially from positive examples with respect to a learned representation are less likely to be synthesizable and are labeled accordingly. Synthesizability classification is then treated as a classic supervised learning problem trained on the now fully labeled dataset. Several related efforts have since followed suit, using PU learning to train synthesizability classifiers with varied training data and model architecture.[30–33] The models use structural and/or compositional information to predict "likelihood scores," indicating the probability that each material can be synthesized (using any synthetic approach). If this score is above a designated threshold (often 0.5), the material is classified as synthesizable.

ML models are appealing for their streamlined and low-cost quantification of synthesizability. However, their utility to synthetic chemists is contingent on their reliability, which is difficult to quantify without access to extensive negative examples (failed synthesis experiments). In the absence of negative data, the thermodynamics of successful syntheses can inform the likelihood of accessing a material. In this work, we evaluate previously published ML model predictions in the context of thermodynamic heuristics. We find that synthesizability predictions from these models do correlate with thermodynamic heuristics to some degree, often assigning lower likelihood scores to materials far above the convex hull. However, the predicted scores remain too high in general, overestimating the synthetic accessibility of materials that are unlikely to be synthesized based on a thermodynamic assessment.

## Results

### Synthesizability assessments from thermodynamic and data-driven approaches

$\Gamma$ quantifies the thermodynamic selectivity of a chosen synthesis recipe (**Equation 1**), whereas ML models for synthesis prediction assess the synthesizability of a chosen material. To align these two approaches, we define a material-centric quantity, $\Gamma_{opt}$, as the lowest (most favorable) $\Gamma$ across all possible precursors for a given target (details in **Methods**). We consider four recently published ML models for synthesizability prediction to compare with $\Gamma_{opt}$, all of which employ some form of PU learning. Of these four models, two use only compositional information (PU-CGNF[30] and SynthNN[31]) while two also use crystal structure information (PU-CGCNN[29] and SynCoTrain[32]). The distinguishing features of each classifier are summarized in **Table 1** and further detailed in the **Methods** section. Given the lack of data on unsynthesizable materials, ML synthesizability predictors are generally assessed on the basis of true positive rate (TPR) or recall. In the context of synthesis prediction, TPR indicates the fraction of synthesized materials that are predicted by the model to be synthesizable. While the specific details associated with training/testing protocols vary across the respective studies, the models employed here report test set TPRs of ~80-90%.

**Table 1.** Distinguishing features of the four ML synthesizability classifiers evaluated in this work. MP refers to the Materials Project[6] and OQMD refers to the Open Quantum Materials Database.[7] SVM refers to support vector machine.

| Model | Material encoding | Source of positive data | Source of unlabeled data | Labeling method |
|---|---|---|---|---|
| **PU-CGNF** | Composition | MP and OQMD known materials | MP and OQMD hypothetical materials | Transductive bagging SVM |
| **SynthNN** | Composition | ICSD | Artificially generated chemical formulae | Probabilistic weighted labeling |
| **PU-CGCNN** | Structure | MP known materials | MP hypothetical materials | Transductive bagging SVM |
| **SynCoTrain** | Structure | MP known oxide materials | MP hypothetical oxide materials | Co-training with alternating iterations of bagging SVMs |

**Calibration on text-mined synthesis data**

A text-mined dataset of successful solid-state synthesis reactions[27] was used to identify the range of Γ consistent with known (successful) reactions. Using DFT-calculated thermodynamic data in MP, each published reaction in the text-mined dataset was mapped to a computed $E_{hull}$ associated with the target material and a computed Γ associated with the experimentally observed recipe ($\Gamma_{obs}$). In **Figure 1a**, $\Gamma_{obs}$ is plotted as a function of the corresponding $E_{hull}$ for 2495 reported reactions targeting ternary metal oxides. Some correlation between $\Gamma_{obs}$ and $E_{hull}$ is observed, with less stable materials often having higher $\Gamma_{obs}$ and therefore synthesis recipes that are predicted to be less selective. However, the mapping between these two quantities is not one-to-one. There is substantial variance in $\Gamma_{obs}$ even at fixed $E_{hull}$, meaning thermodynamic selectivity can differ substantially for two materials that are comparably close to the convex hull, supporting the notion that these metrics provide complementary information.

In **Figure 1b**, we illustrate how Γ can vary dramatically at relatively constant $E_{hull}$ by comparing $LaNb_3O_9$ ($E_{hull}$ = 31 meV/atom, $\Gamma_{obs}$ = 102 meV/atom) and $LaGaO_3$ ($E_{hull}$ = 33 meV/atom, $\Gamma_{obs}$ = 40 meV/atom). The much larger Γ for $LaNb_3O_9$ can be understood by examining the interface reaction hulls in **Figure 1b**, which represent syntheses listed in the text-mined dataset. Whereas the two products that compete with $LaGaO_3$ (right-hand hull, blue) are comparably "shallow" on the hull – meaning they have small reaction energies to form – the products that compete with $LaNb_3O_9$ lie much deeper on the interface hull. As a result, primary competition for $LaNb_3O_9$ formation is high ($C_1$ = 115 meV/atom) since the most competitive product ($LaNbO_4$) has a substantially greater driving force to form from $Nb_2O_5$ and $La_2O_3$. Secondary competition is similarly quite high ($C_2$ = 186 meV/atom) since it is highly favorable for the target to react with remaining precursors to form $LaNbO_4$ and others on the $La_2O_3$-rich side of the hull. Meanwhile, the reaction energy to form $LaGaO_3$ is comparable to competing targets, resulting in much lower primary competition ($C_1$ = 57 meV/atom). The driving forces associated with $LaGaO_3$ reactions with either precursor are also lower than in the La-Nb-O case, leading to lower secondary competition ($C_2$ = 55 meV/atom). Despite the similar $E_{hull}$ for the two targets, the choice of precursors and distribution of competing targets in each chemical space leads to distinct Γ, demonstrating that Γ offers additional insight beyond $E_{hull}$ when screening for synthesizability. Note also in **Figure 1b** that the two targets are different distances from the interface hull at their reported synthesis temperatures. Although they have very similar $E_{hull}$ at 0 K, $LaNb_3O_9$ stabilizes and $LaGaO_3$ destabilizes with increasing temperature. This behavior further exemplifies the distinction between stability and selectivity. $LaNb_3O_9$ is much closer to the hull at its listed reaction temperature than $LaGaO_3$ but is still a less favorable target from a selectivity perspective due to the presence of more thermodynamically competitive compounds in the La-Nb-O chemical space.

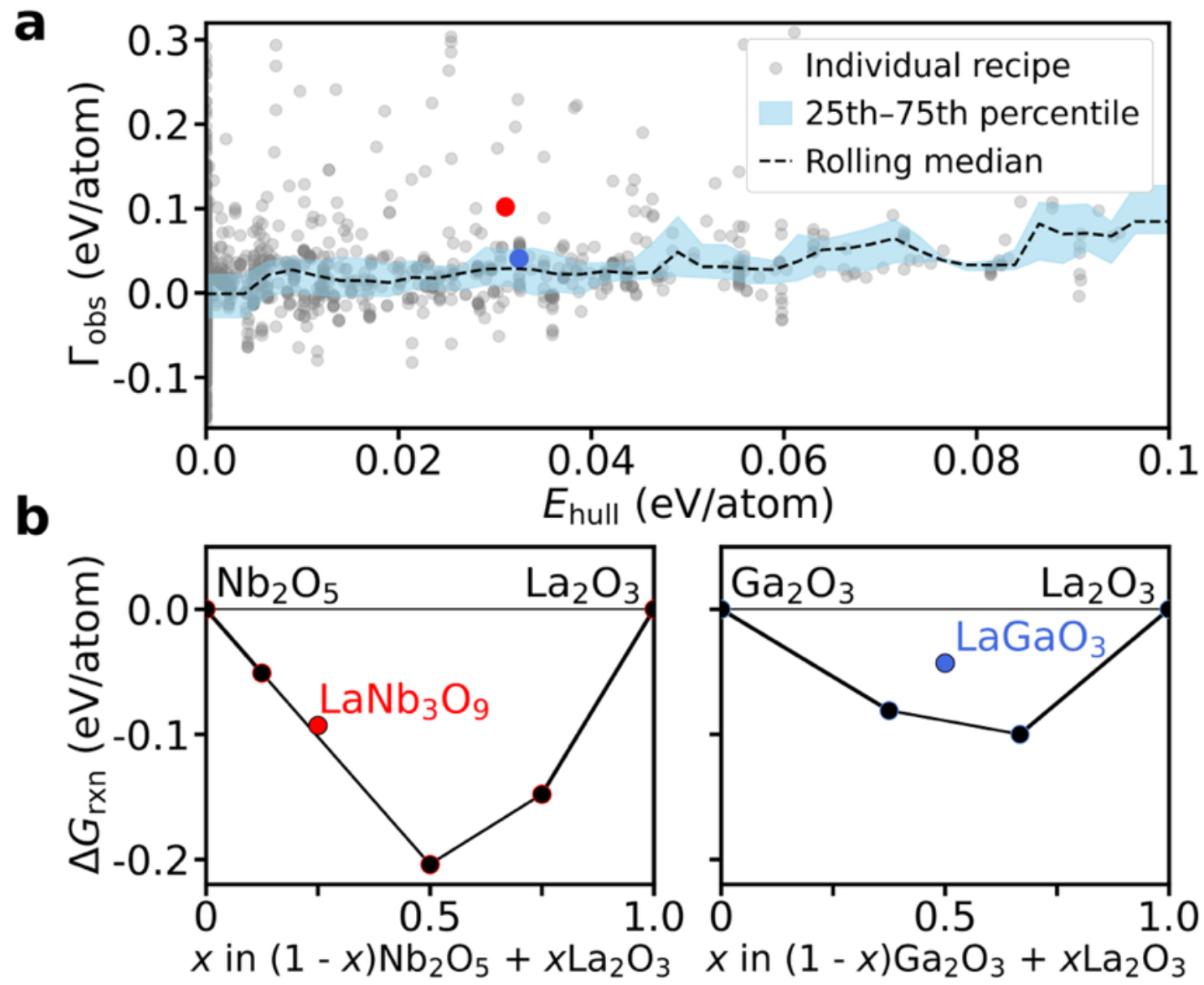

**Figure 1. (a)** $\Gamma_{obs}$ vs. target $E_{hull}$ for each reaction listed in the text-mined dataset (grey points) and a rolling median of $\Gamma_{obs}$ as a function of target $E_{hull}$, with the median evaluated over a range of 8 meV/atom every 2.5 meV/atom. The distance from the median to the 25$^{th}$ and 75$^{th}$ percentile values of $\Gamma_{obs}$ is shaded in light blue. The examples shown in (b) are indicated with red and blue points. $\Gamma_{obs}$ was calculated at the reported synthesis temperature. $E_{hull}$ was calculated at 0 K. **(b)** Two interface reaction hulls depicting the relative driving forces to form different products from a given precursor pair. The left-hand hull represents the reaction between $Nb_2O_5$ and $La_2O_3$ targeting $LaNb_3O_9$, and the right-hand hull represents the reaction between $Ga_2O_3$ and $La_2O_3$ targeting $LaGaO_3$. These interface hulls are presented at the observed synthesis temperatures of 1273 K and 723 K for $LaNb_3O_9$[34] and $LaGaO_3$,[35] respectively.

The reaction-centric $\Gamma_{obs}$ for each text-mined entry is also compared to the material-centric $\Gamma_{opt}$ associated with the listed target to validate the efficacy of $\Gamma_{opt}$ as a proxy for $\Gamma_{obs}$. This becomes important for hypothetical materials where $\Gamma_{obs}$ is unknown (as no synthesis has been reported). A hexbin parity plot comparing $\Gamma_{obs}$ to $\Gamma_{opt}$ for the 2495 listed reactions (spanning 938 unique targets) is shown in **Figure 2a**. The highest density of points lies near the parity line, indicating that observed (successful) reactions often approach the optimal (minimum) selectivity. In fact, 41% of the experimentally reported reactions in the text-mined dataset use the exact precursors that yield $\Gamma_{opt}$, and 63% of the experimentally reported reactions have $\Gamma_{obs}$ within 20 meV/atom of $\Gamma_{opt}$. These results indicate that successful synthesis routes are typically associated with selectivities close to the optimum, supporting the use of $\Gamma_{opt}$ as a practical thermodynamic heuristic for quantifying the synthesizability of hypothetical materials.

In **Figure 2b**, we show cumulative distribution functions for both observed reactions ($\Gamma_{obs}$) and their corresponding optimal values ($\Gamma_{opt}$), showing the fraction of reactions ($\Gamma_{obs}$) and targets ($\Gamma_{opt}$) with $\Gamma$ below a given threshold. The closeness of the two distributions further underscores the similarity of $\Gamma_{obs}$ and $\Gamma_{opt}$. Examining the distribution for $\Gamma_{obs}$, 92% of observed (successful) reactions have $\Gamma_{obs} < 100$ meV/atom, as demarcated by the dashed vertical line. Materials with $\Gamma_{opt} > 100$ meV/atom are therefore considered unlikely to be synthesizable, since even their most selective pathways fall outside the bounds of the vast majority of this set of reactions. Applying similar logic to $E_{hull}$, the aforementioned study by Sun et al. found that ~95% of materials listed in the ICSD have $E_{hull} < 67$ meV/atom.[17] Allowing for some outliers, materials with $E_{hull} > 100$ meV/atom are similarly considered unlikely to be synthesizable. These results suggest that hypothetical materials having either $\Gamma_{opt} > 100$ meV/atom or $E_{hull} > 100$ meV/atom are unlikely to be synthesized, rendering these useful metrics for assessing ML models in the absence of explicit negative examples. While these proposed bounds are informed by a statistical analysis of known materials and their synthesis recipes, they should not be interpreted as strict cutoffs. It is likely that some materials within these bounds ($\Gamma_{opt} < 100$ meV/atom and $E_{hull} < 100$ meV/atom) cannot be synthesized and some materials outside these bounds ($\Gamma_{opt} > 100$ meV/atom or $E_{hull} > 100$ meV/atom) can be synthesized. However, given the analysis of known materials, we should expect the likelihood of synthesis to decrease as these thermodynamic heuristics increase and approach zero as these thermodynamic heuristics increase significantly beyond these bounds (e.g., $E_{hull}$ or $\Gamma_{opt} >> 0.1$ eV/atom).

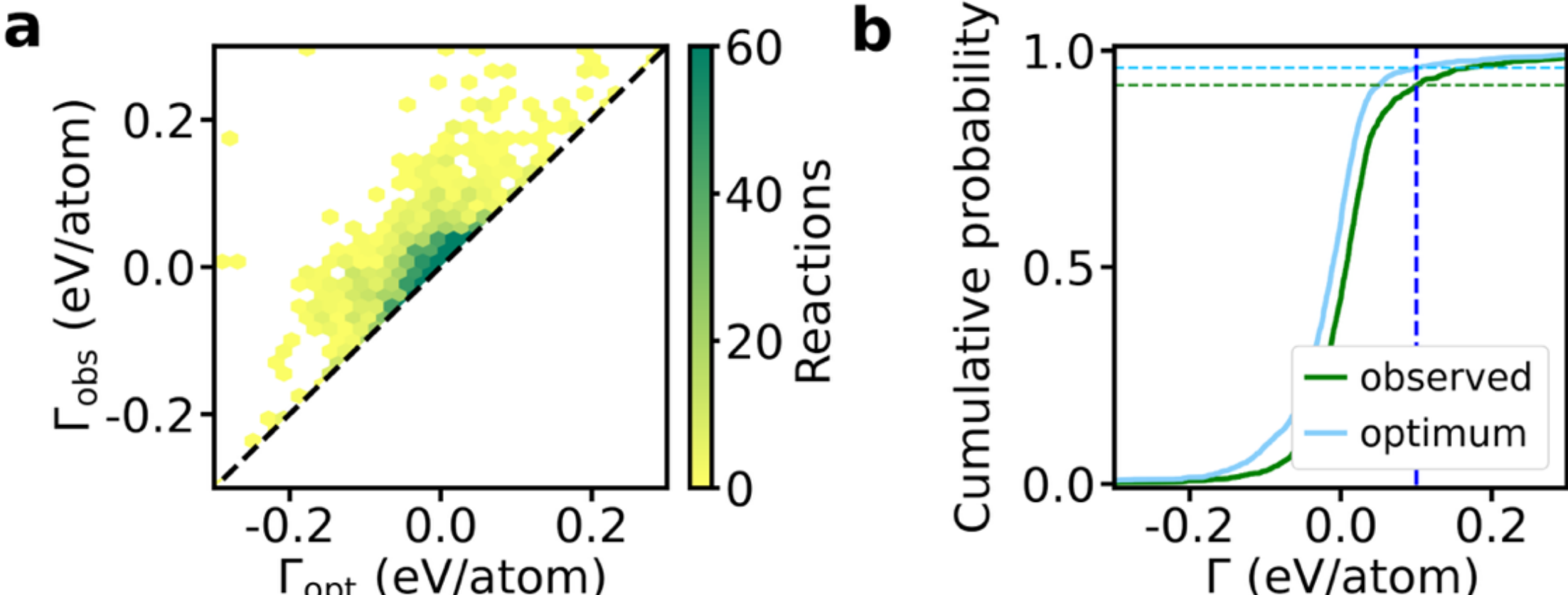

**Figure 2. (a)** A pairwise comparison of $\Gamma$ for 2495 observed reactions[27] ($\Gamma_{obs}$) with the optimum value ($\Gamma_{opt}$) that is possible from known precursors to make the same target. Bins are colored by the number of reactions falling within each region. **(b)** Cumulative distributions of $\Gamma_{obs}$ (green) and $\Gamma_{opt}$ (light blue) for text-mined reactions. The vertical blue dashed line indicates $\Gamma = 100$ meV/atom, corresponding to the 92$^{nd}$ percentile of $\Gamma_{obs}$ (green dashed horizontal line) and the 96$^{th}$ percentile of $\Gamma_{opt}$ (light blue dashed horizontal line). Tabulated values corresponding to these plots are available in **Table S1**.

### Data-driven versus thermodynamic predictions on hypothetical materials

Using the bounds on $\Gamma_{opt}$ established with the text-mined synthesis recipes, we sought to assess ML synthesizability predictors on a new dataset of hypothetical materials. This dataset is comprised of 2673 hypothetical materials generated using the generative AI model, Chemeleon.[36] These materials span the same ternary oxide chemical spaces included in the text-mined dataset but do not appear in the ICSD or MP, meaning they were unlikely to have been used during the training of any models studied here. Further details on the generation procedure are provided in **Methods**. Each generated structure was relaxed using CHGNet,[11] and the resulting energies were used to estimate $E_{hull}$ and $\Gamma_{opt}$ in the same manner as for the text-mined data (leveraging MP data for other materials in each relevant chemical space).

The resulting $E_{hull}$ distribution for the 2673 generated materials is shown in **Figure 3a**. The majority of generated materials (~60%) have $E_{hull}$ > 100 meV/atom, meaning they are unlikely to be synthesizable based on a stability assessment. For the remaining 1057 materials that are stable or plausibly metastable ($E_{hull}$ < 100 meV/atom), we show the distribution of $\Gamma_{opt}$ in **Figure 3b**. Even for these materials that cannot be ruled out from a stability perspective, ~60% have thermodynamic selectivities ($\Gamma_{opt}$) greater than 100 meV/atom, falling outside the range where successful synthesis is expected to be likely. This dataset presents a wide spectrum of new hypothetical materials where synthesis appears plausible (low $E_{hull}$ and low $\Gamma_{opt}$), and importantly, many examples where ML models should assign low likelihood scores (high $E_{hull}$ or high $\Gamma_{opt}$). It should be noted that previous work found CHGNet to slightly overestimate the stability (underestimate $E_{hull}$) of materials generated by a variety of generative models, suggesting the thermodynamic predictions reported here may themselves overestimate the synthesizability of these hypothetical materials.[37]

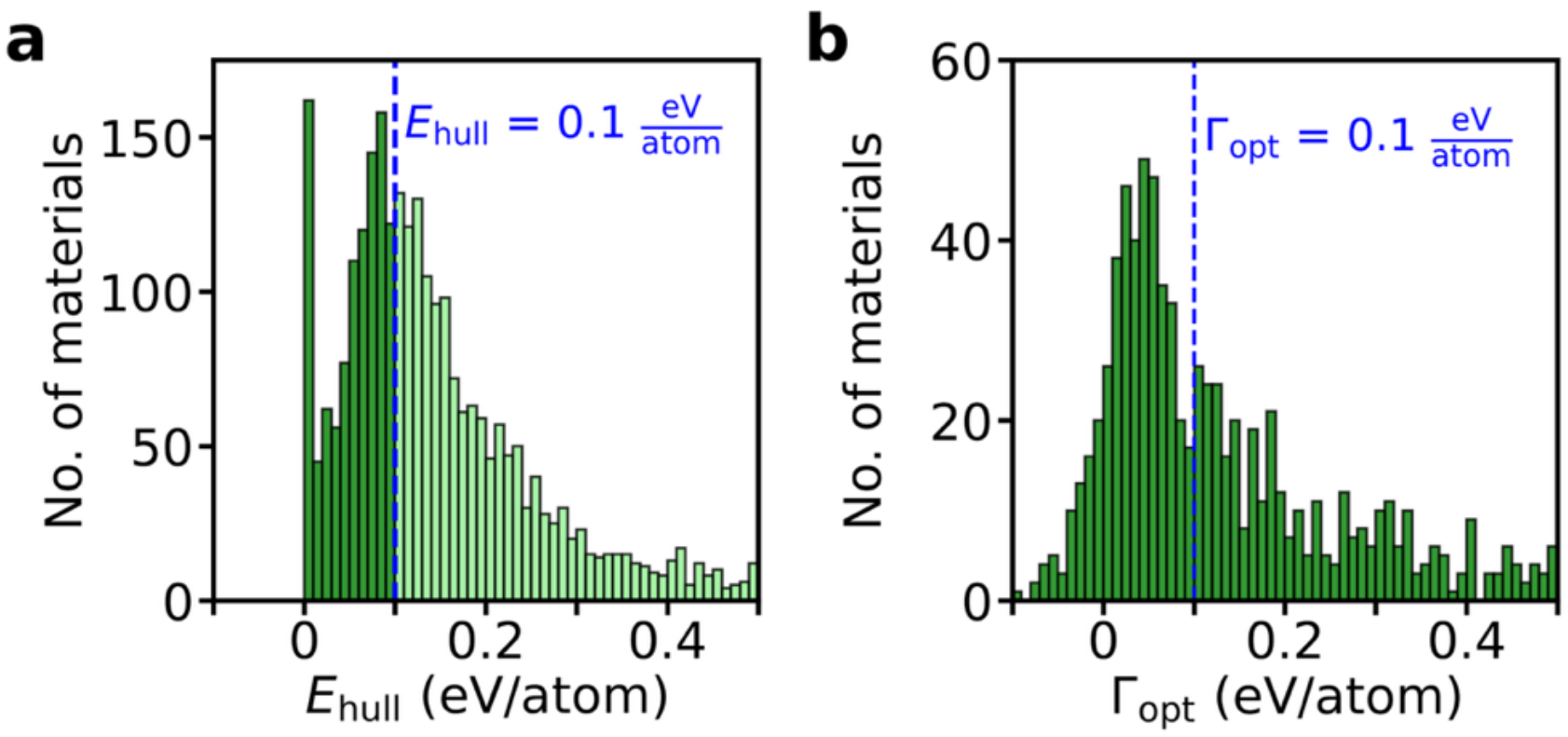

**Figure 3.** Distributions of $E_{hull}$ **(a)** and $\Gamma_{opt}$ **(b)** for Chemeleon-generated materials, with proposed bounds on thermodynamic feasibility ($E_{hull}$ < 100 meV/atom and $\Gamma_{opt}$ < 100 meV/atom), demarcated by the blue dashed lines. The $\Gamma_{opt}$ distribution reflects only materials within the dark green bars in (a) – those with $E_{hull}$ < 100 meV/atom. The energies of generated materials were computed with CHGNet.[11] The calculation of $E_{hull}$ and $\Gamma_{opt}$ from these energies are detailed further in **Methods**.

The four aforementioned synthesizability predictors (**Table 1**) were applied to these newly generated materials. In **Table 2**, we list the number of hypothetical materials predicted synthesizable by each model and the number of materials within specified thermodynamic bounds. While 40% of materials meet our bound on thermodynamic stability ($E_{\mathrm{hull}} < 100$ meV/atom), only 16% of materials fall within both proposed thermodynamic bounds ($E_{\mathrm{hull}} < 100$ meV/atom and $\Gamma_{\mathrm{opt}} < 100$ meV/atom), indicating the value of thermodynamic selectivity metrics in decreasing the breadth of hypothetical materials that should be targeted for synthesis. All ML models except SynthNN predict at least 50% of generated materials to be synthesizable, indicating a propensity for overprediction. It is notable that the fraction of materials predicted synthesizable by SynthNN (14%) is similar to the fraction that look promising based on thermodynamic heuristics (16%).

**Table 2.** Number and fraction of Chemeleon-generated hypothetical materials predicted to be synthesizable by four ML classifiers, along with subsets of materials that fall within realistic thermodynamic bounds.

| | # of materials | % of materials |
|---|---:|---:|
| Total hypothetical materials considered | 2673 | N/A |
| Predicted synthesizable (PU-CGNF) | 2050 | 77% |
| Predicted synthesizable (SynthNN) | 377 | 14% |
| Predicted synthesizable (PU-CGCNN) | 1326 | 50% |
| Predicted synthesizable (SynCoTrain) | 1929 | 72% |
| $E_{\mathrm{hull}} < 100$ meV/atom | 1057 | 40% |
| $E_{\mathrm{hull}} < 100$ meV/atom and $\Gamma_{\mathrm{opt}} < 100$ meV/atom | 427 | 16% |

Beyond aggregated statistics, synthesizability scores predicted by each model for each material were compared to their thermodynamic heuristics, $E_{\mathrm{hull}}$ and $\Gamma_{\mathrm{opt}}$. In **Figure 4**, we show heatmap distributions of synthesizability scores as a function of $E_{\mathrm{hull}}$ for each model. Dark green regions indicate a high density of materials with a given $E_{\mathrm{hull}}$ and synthesizability score. Overlain on each heatmap is a moving average of the fraction of materials predicted synthesizable as a function of $E_{\mathrm{hull}}$ (indicated on the left-hand $y$-axis).

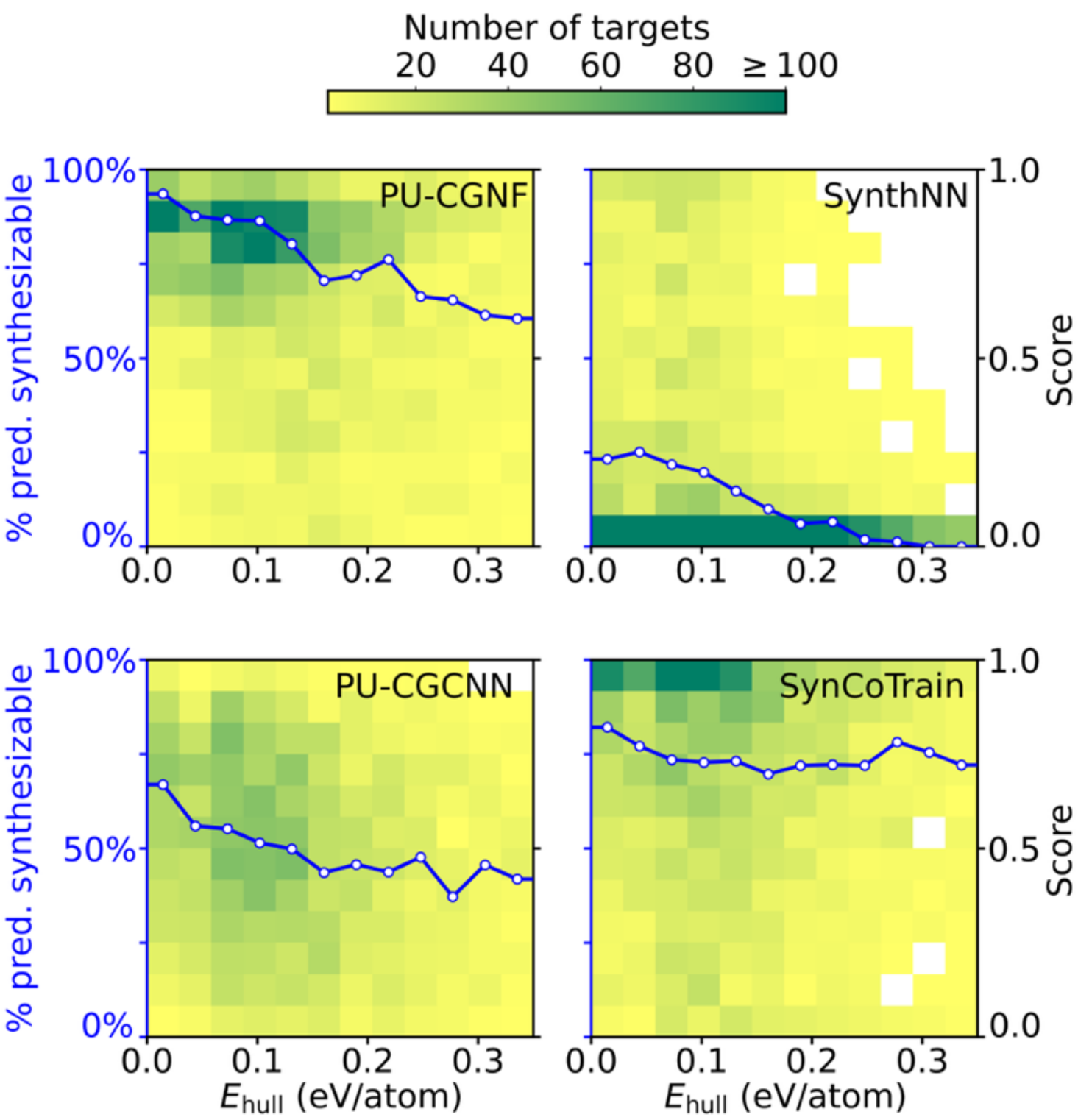

**Figure 4.** Model predictions and CHGNet-calculated $E_{\mathrm{hull}}$ for 2673 Chemeleon-generated materials. The heatmap color indicates the number of these target materials having a certain synthesizability score (right-hand *y*-axis) and $E_{\mathrm{hull}}$ (*x*-axis). A moving average of the fraction of materials predicted synthesizable (score > 0.5) as a function of $E_{\mathrm{hull}}$ is shown by the blue line and corresponding left-hand *y*-axis. The moving average was computed over 30 meV/atom intervals. White squares indicate no targets falling in a given region. The model name is given in the top right of each subplot. Tabulated values corresponding to these plots are available in **Table S2**.

Stable ($E_{\mathrm{hull}} = 0$) and weakly unstable materials ($E_{\mathrm{hull}} < 100$ meV/atom) are considered here to be feasible candidates for synthesis, given the distribution of $E_{\mathrm{hull}}$ determined by Sun *et al.*[17] for known materials. In contrast, high-$E_{\mathrm{hull}}$ materials are unlikely to be synthesized by traditional approaches such as solid-state synthesis. A predictor that successfully aligns with thermodynamic trends should therefore yield a relatively high density of scores > 0.5 for materials with $E_{\mathrm{hull}} \ll$ 100 meV/atom, and progressively lower scores for materials far above the hull. From **Figure 4**, it is clear that each of the models predicts stable materials to be synthesizable with the greatest frequency, as expected. However, even for stable materials ($E_{\mathrm{hull}} = 0$), there is a wide range of outcomes depending on the model. For example, of the 136 stable materials, SynthNN predicts 25% to be synthesizable whereas PU-CGNF predicts 93% to be synthesizable. As materials get less stable (larger $E_{\mathrm{hull}}$), models should predict exponentially fewer can be synthesized. While there is a general decreasing trend in the synthesizability scores from PU-CGNF, SynthNN, and PU-CGCNN, SynCoTrain predictions remain relatively constant regardless of $E_{\mathrm{hull}}$. As $E_{\mathrm{hull}}$ increases beyond 100 meV/atom, we would expect all models to rapidly decrease in predicted synthesizability score, but only SynthNN predictions go to zero even for the most unstable materials ($E_{hull}$ >> 100 meV/atom). Among the 679 generated materials with $E_{hull}$ > 200 meV/atom, PU-CGCNN, PU-CGNF, and SynCoTrain predict 39%, 56%, and 66% to be synthesizable, respectively, but the synthesis of any of these highly unstable materials would be exceptional.

Among generated materials with $E_{hull}$ < 100 meV/atom, synthesis is plausible but may not be straightforward. Here, $\Gamma_{opt}$ is used as an additional thermodynamic heuristic to probe the synthesizability of materials in the low-$E_{hull}$ regime. In **Figure 5**, we show heatmap distributions of synthesizability score as a function of $\Gamma_{opt}$ for each model. Only the subset of materials with $E_{hull}$ < 100 meV/atom are considered here. Dark green regions indicate a high density of materials with a given $\Gamma_{opt}$ and synthesizability score. Overlain on the heatmaps are moving averages of the fraction of materials predicted synthesizable (score > 0.5) by each model as a function of $\Gamma_{opt}$.

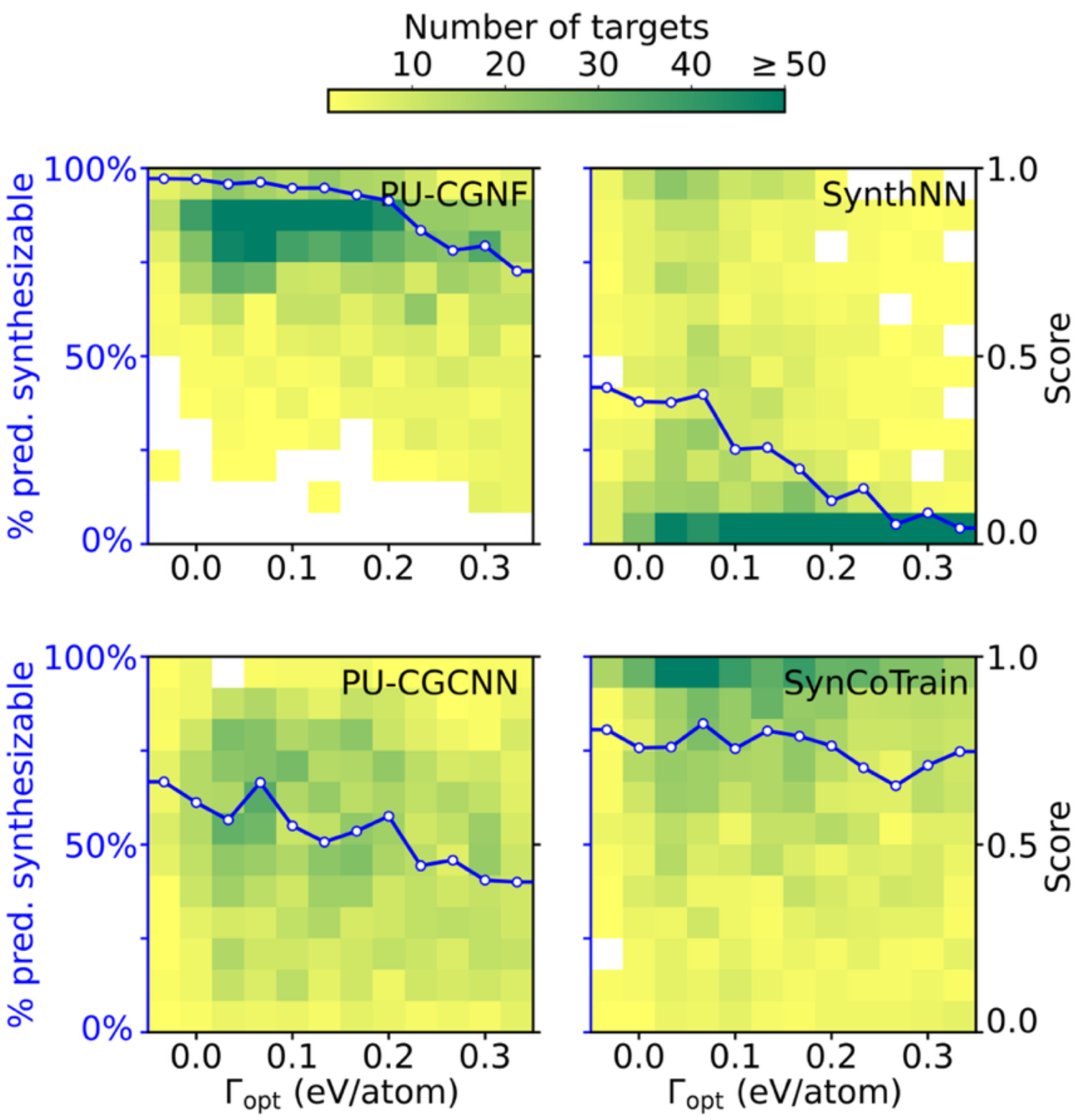

**Figure 5.** Model predictions and CHGNet-calculated $\Gamma_{opt}$ for 1057 Chemeleon-generated materials having $E_{hull}$ < 100 meV/atom. The heatmap color indicates the number of these target materials having a certain synthesizability score (right-hand *y*-axis) and $\Gamma_{opt}$ (*x*-axis). A moving average of the fraction of materials predicted synthesizable (score > 0.5) as a function of $\Gamma_{opt}$ is shown by the blue line and corresponding left-hand *y*-axis. The moving average was computed over 33 meV/atom intervals. White squares indicate no targets falling in a given region. The model name is given in the top right of each subplot. Tabulated values corresponding to these plots are available in **Table S2**.

Even for materials with low $E_{hull}$, the likelihood of synthesis should decrease with increasing $\Gamma_{opt}$ since a high (positive) $\Gamma_{opt}$ indicates an absence of selective pathways to the target in question. Recall that just 40% of generated materials with $E_{hull}$ < 100 meV/atom have associated $\Gamma_{opt}$ < 100 meV/atom. However, PU-CGNF, PU-CGCNN, and SynCoTrain predict 89%, 58%, and 77% of this low-$E_{hull}$ subset to be synthesizable, indicating that synthesizability is overestimated by all models other than SynthNN (23%) with respect to $\Gamma_{opt}$ as well as $E_{hull}$. Remarkably, as with $E_{hull}$, SynthNN shows a marked decrease in synthesizability score as $\Gamma_{opt}$ increases. While 23% of $E_{hull}$ < 100 meV/atom materials are predicted synthesizable by SynthNN, this fraction increases to 40% if $\Gamma_{opt}$ < 100 meV/atom and decreases to 18% if $\Gamma_{opt}$ > 100 meV/atom. This shift cannot be explained solely by variation in $E_{hull}$ within the smaller range $E_{hull}$ < 100 meV/atom as $\Gamma_{opt}$ and $E_{hull}$ are only weakly correlated over this subset of generated materials (**Figure S1**). In contrast, PU-CGNF, PU-CGCNN, and SynCoTrain exhibit weak downward trends with increasing $\Gamma_{opt}$ or general insensitivity to $\Gamma_{opt}$ for materials in this range of $E_{hull}$.

**Comparison among ML models**

The synthesizability predictors were also examined relative to one another to understand whether they are learning similar underlying information about the generated materials. The histograms in **Figure 6a** show the distribution of synthesizability scores associated with the generated materials for each model. As indicated by the summary statistics in **Table 2**, SynthNN is heavily skewed toward classifying materials as unsynthesizable, while PU-CGNF and SynCoTrain are inclined to classify materials as synthesizable. PU-CGCNN predictions are distributed more evenly, with an inclination towards low-certainty scores (~0.5). While the "true" distribution is unknown since we considered only hypothetical (not-yet-synthesized) materials, SynthNN appears most true to the thermodynamic distributions shown in **Figure 3**. However, SynthNN predicts substantially fewer materials to be synthesizable than the number that do reside within the bounds, suggesting it may be overly pessimistic in some cases.

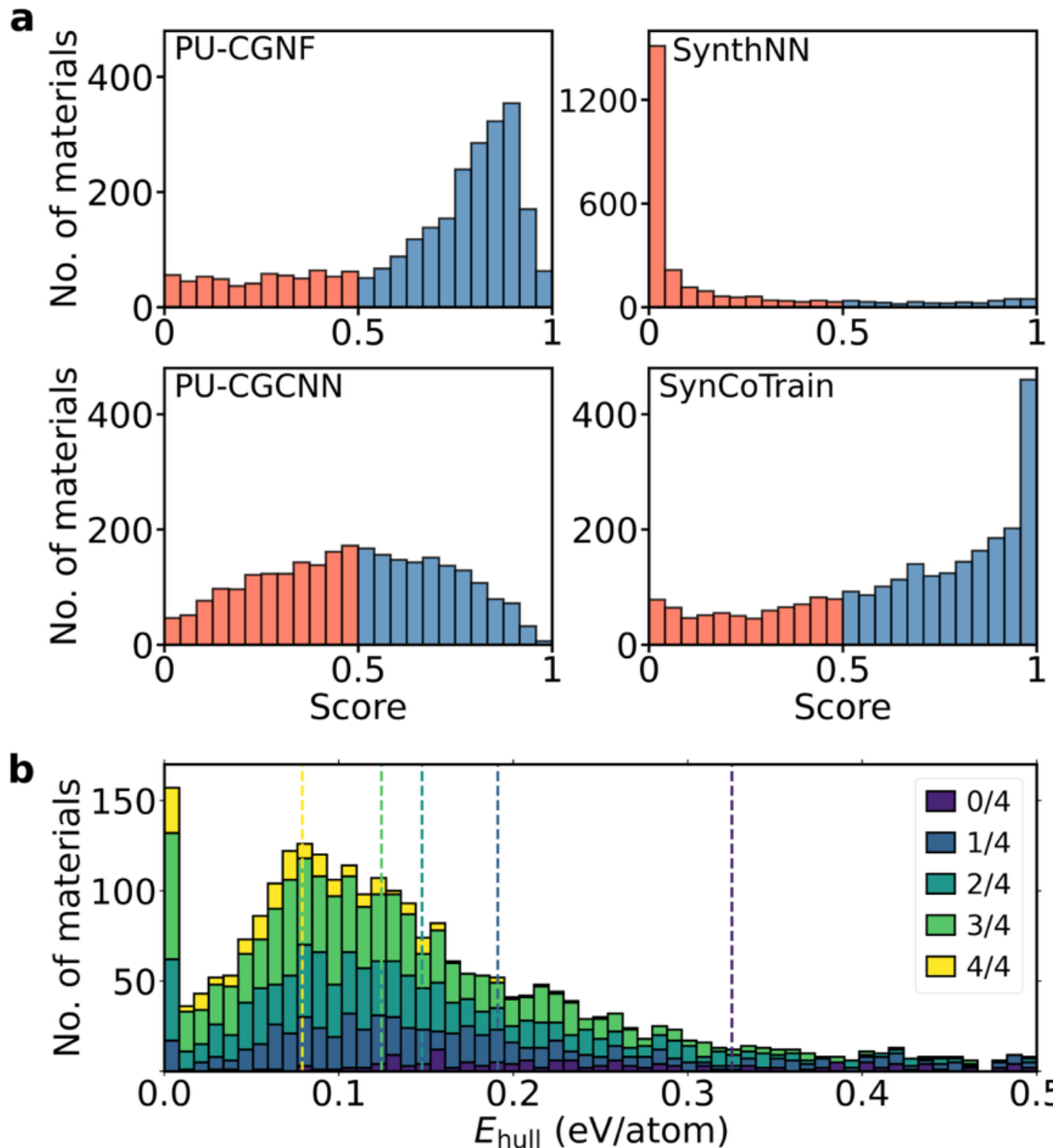

**Figure 6. (a)** Distributions of predicted synthesizability scores from each model for the 2673 Chemeleon-generated materials. Scores corresponding to a synthesizable label are colored in blue (> 0.5) while those corresponding to an unsynthesizable label (< 0.5) are colored in red. Note that the *y*-axis of SynthNN (upper right) is plotted on a different scale than the other models due to the dramatic skew of scores toward low values. **(b)** A stacked histogram of materials predicted synthesizable by a specified fraction of the models as a function of $E_{hull}$. For example, "4/4" indicates that all four models (PU-CGNF, SynthNN, PU-CGCNN, and SynCoTrain) predict a material to be synthesizable. The dashed lines in the corresponding colors indicate the mean $E_{hull}$ of materials predicted synthesizable by that fraction of models.

In **Figure 6b**, model agreement is assessed by indicating how many models predict each material to be synthesizable as a function of $E_{hull}$. Within a given range of $E_{hull}$, the number of materials classified synthesizable by a specified fraction of the models (from 0/4 to 4/4) is indicated by the height of the bar with the associated color. The distributions corresponding to materials classified as synthesizable by 0/4 models and 4/4 models (purple and yellow bars, respectively) indicate universal agreement among models. These are the least prominent distributions with just 7% of materials being predicted synthesizable and 7% predicted unsynthesizable by all four models. The other distributions (1/4, 2/4, 3/4) all indicate disagreement by at least one model. The most common occurrence, indicated by the bars with the greatest height across $E_{hull}$, is of materials predicted synthesizable by 3/4 models, with SynthNN being the lone model classifying the material as unsynthesizable in 87% of these instances. Materials predicted synthesizable by all four models have the lowest mean hull energy (79 meV/atom), as indicated by the yellow dashed line. This reflects the propensity of SynthNN to classify only lower $E_{hull}$ materials as synthesizable as well as the decreasing number of materials classified as synthesizable by PU-CGNF and PU-CGCNN with increasing $E_{hull}$ (**Figure 4**). Meanwhile, PU-CGNF and SynCoTrain together comprise 90% of cases where a material is predicted synthesizable by only one model. However, the mean $E_{hull}$ is 70 meV/atom higher for materials predicted synthesizable only by SynCoTrain than for materials predicted synthesizable only by PU-CGNF. This reflects the propensity of PU-CGNF to classify fewer materials as synthesizable with increasing $E_{hull}$, while SynCoTrain scores do not vary substantially with $E_{hull}$.

While these distributions paint an aggregated picture, comparing the models on a target basis shows that model scores vary substantially for individual materials. A pairwise plot of scores for each pair of models is shown in **Figure S2**. The greatest correlation exists among PU-CGNF and SynCoTrain, both having a general propensity for positive classification. The general lack of correlation between these models indicates significant sensitivity with respect to model choices such as training data, labeling strategy, and architecture. If each model's predictions are reliable, it is possible that ensembling may yield complementary information. However, the general disagreement between model predictions and thermodynamic trends does not support this being a viable strategy with current models.

## Discussion

ML models have become reliable tools for many materials science problems, including thermodynamic stability prediction.[38] However, the task of predictive synthesis is more challenging due to the complexity of phase transformations and relative lack of available data for training. Among other challenges, the efficacy of these models cannot be fully quantified in the absence of known unsynthesizable materials (negative examples). In this work, we bypass this data shortage by using thermodynamic heuristics ($E_{hull}$, $\Gamma_{opt}$) to identify which materials are more or less likely to be synthesizable. Sun et al. reported 95% of synthesized materials in the ICSD exhibit $E_{hull} \leq 67$ meV/atom).[17] In this work, we find that 96% of studied reactions exhibit $\Gamma_{opt} \leq$ 100 meV/atom. Although particular values of these thermodynamic quantities cannot guarantee (un)synthesizability, they reflect the probability distributions associated with the known materials and syntheses in this analysis. As either $E_{hull}$ or $\Gamma_{opt}$ increase, these findings suggest that synthesis should become less likely.

Four ML models for synthesizability prediction – PU-CGNF[30], SynthNN[31], PU-CGCNN[29], and SynCoTrain[32] – were evaluated against these computed thermodynamic heuristics. Three of the four models systematically overestimate synthesizability compared with the bounds on thermodynamic stability and selectivity established on known materials. SynthNN was the only model with drastically decreasing synthesizability scores as $E_{hull}$ increases beyond 100 meV/atom (**Figure 4**). For materials with plausible thermodynamic stability ($E_{hull} < 100$ meV/atom), thermodynamic selectivity as assessed by $\Gamma_{opt}$ was shown to further distinguish between likely and unlikely candidates for synthesis. Again, SynthNN was the only model to display a strong trend towards unsynthesizable predictions as $\Gamma_{opt}$ increased (**Figure 5**).

The relative success of SynthNN might be explained by the choice of unlabeled materials during training. While all other models use hypothetical materials from materials databases (MP or OQMD) as unlabeled materials, SynthNN generates entirely new hypothetical materials (chemical formulas) spanning a similar distribution to the known materials in the ICSD. While the provenance of hypothetical materials in materials databases varies, these materials may exhibit greater similarity to known materials than the artificially generated formulas in the SynthNN work. Models trained on materials databases may therefore be biased towards classifying more materials as synthesizable. SynthNN may also benefit from a much larger quantity of unlabeled data relative to the other models since the unlabeled data is not sourced from existing databases. Our results indicate that SynthNN exhibits unique behavior among the four models studied, and future work could isolate the effect of training data on this finding.

In addition, PU-CGNF, PU-CGCNN, and SynCoTrain employ hard labeling, which treats borderline-class materials (scores ≈ 0.5) the same as materials with very high (≈1) and very low (≈0) scores. In contrast, SynthNN privileges materials with scores close to 0 or 1 while down-weighting more ambiguous unlabeled materials in training. This labeling strategy targets precision (the fraction of labeled positives that are truly positive) rather than recall (the fraction of positives correctly identified as positive). As a result, positive labels are assigned more conservatively, resulting in generally lower positive class probabilities. Exposure to more unrealistic materials along with confidence-based labeling may be responsible for better alignment between SynthNN predictions and computed thermodynamic heuristics.

The models considered here are distinguished not only by training data and labeling strategy, but also by material representation. To determine the effect of material representation on model performance, two structural models, PU-CGCNN and SynCoTrain, were compared to two compositional models, SynthNN and PU-CGNF. Prior deep learning models have found substantial improvement with structural representations over compositional representations, which is not surprising given the ubiquity of structure-property relations in materials science.[10] However, the results in this work indicate greater ambiguity for the problem of synthesizability prediction as a compositional model (SynthNN) exhibits the greatest alignment with thermodynamic heuristics. Among the other models, it is difficult to discern a notable benefit (or detriment) of including structure in the representation. A third structural model, TSDNN[33] was examined for further insight on representation effects. TSDNN predicted 98% of the generated materials to be synthesizable, with little variation in score as a function of $E_{hull}$ or $\Gamma_{opt}$, leading to the likely erroneous prediction of many thermodynamically unrealistic materials as synthesizable (see **Figure S3**). The limited performance gains when including structure in the representation stand in stark contrast to ML models that predict other material properties (e.g., formation energy) and may arise from the relatively smaller datasets available for synthesizability prediction or the disconnect between structural features and synthesis.

The general challenge of predicting synthesizability also stems from the inherently non-intrinsic nature of "synthesis". Throughout this work, synthesizability is framed as a material property with materials being labeled as either synthesizable or unsynthesizable. In reality, whether or not a material can be synthesized depends critically on the synthesis recipe (precursors, temperature, approach, etc.).[15,19,39] From an ML perspective, this presents a challenge compared with the prediction of calculable, intrinsic material properties (formation energy, band gap, etc.). Thermodynamic heuristics that link materials to the plausibility of synthesis are useful in this regard, with $E_{\mathrm{hull}}$ being the most widely applied and well understood. Enumerating reaction networks and computing thermodynamic selectivities associated with synthesis recipes adds complementary information to $E_{\mathrm{hull}}$ by exploring the landscape of possible synthetic routes to access stable or weakly unstable materials.

While useful, these thermodynamic heuristics are also limited predictors of synthesizability. $E_{\mathrm{hull}}$ generally describes the 0 K bulk energetics of pristine materials at equilibrium. In contrast, materials are synthesized at finite temperatures, likely contain defects, and may not reach equilibrium. While $\Gamma_{\mathrm{opt}}$ accounts for some aspects of possible synthesis recipes, this metric is also limited to bulk thermodynamics, neglecting relevant kinetic processes like diffusion and nucleation as well as the thermodynamics of interfaces (where reactions begin). In addition, both $\Gamma_{\mathrm{opt}}$ and $E_{\mathrm{hull}}$ are limited to the completeness of computed databases. This could lead to an underestimation of both $\Gamma_{\mathrm{opt}}$ and $E_{\mathrm{hull}}$ if important competing phases are not yet tabulated in these databases. Alternatively, $\Gamma_{\mathrm{opt}}$ may be overestimated if certain precursors that are not yet tabulated could lead to thermodynamically selective routes to a given target. In addition, the existing implementation of $\Gamma_{\mathrm{opt}}$ is limited to synthesis recipes with just two solid precursors and no solid byproducts, but more creative recipes (e.g., metathesis) may improve the selectivity.[40] Further, the methods employed in this work focus on the internal energy (as computed using DFT or CHGNet) and a model for the vibrational entropy of solids,[41] so other contributions to the Gibbs energy (e.g., configurational entropy) are neglected. For the targeted synthesis of materials where other effects are important and calculable with first-principles methods (e.g., configurational entropy of high-entropy materials), these quantities can be included in the expressions for the Gibbs energy of each relevant compound before computing $E_{\mathrm{hull}}$ or $\Gamma_{\mathrm{opt}}$. $E_{\mathrm{hull}}$ and $\Gamma_{\mathrm{opt}}$ as presented here therefore function best as preliminary and general screening tools and do not definitively indicate that a material is or is not synthesizable. The synthetic approach is also an important consideration as materials that cannot be made by solid-state synthesis may be synthesizable through other routes (e.g., solution-based synthesis).[28,42] As the thermodynamic and kinetic factors dictating reaction pathways differ as the synthetic approach is varied, it is important to consider how the chosen thermodynamic heuristics may or may not apply across synthetic approaches.[43]

Given the multi-faceted challenge of making new materials, ML synthesizability predictors present an appealing tool for candidate screening as predictions could be made on thousands of materials nearly instantly once these models are trained. Ideally, these models should reflect the kinetic and thermodynamic interplay underpinning synthesis. At a minimum, they should reflect our current understanding of the relationship between synthetic plausibility and thermodynamic quantities. While thermodynamic bounds are not a guarantee of (un)synthesizability, they serve as a strong indicator of likely or unlikely candidates for synthesis. They are leveraged here to identify qualitative and quantitative trends with ML synthesizability predictors as a means of evaluating the performance of these models. We find that the ML models in question produce scores that often overestimate the likelihood that materials can be synthesized compared to established thermodynamic heuristics. Moving forward, we speculate that directly incorporating thermodynamic information into the training and validation process has the potential to improve the ability for models to capture the underlying thermodynamics. The relative success of SynthNN in this regard indicates that training on a wide variety of unlabeled materials and a soft labeling approach may be effective strategies to build on for synthesizability prediction. Ultimately, the efficacy of any synthesis prediction model needs to be validated with experimental synthesis attempts. Currently, synthesizability prediction models likely do not exceed the utility of computable thermodynamic heuristics. While synthesizability predictions are nearly free from a computational standpoint (once models are trained), the time required to perform first-principles estimates of stability and selectivity (~hours per material) remains far lower than the time required to perform experiments (~days per material). Replacing DFT with foundation potentials further reduces the time required to compute thermodynamic heuristics to ~minutes per material.

As generative models rapidly expand the space of proposed materials and large-scale libraries of hypothetical compounds continue to grow, the need for accurate and efficient synthesis prediction has never been greater. Along with calculable thermodynamic quantities, the increasing prevalence of self-driving labs for inorganic synthesis presents an opportunity to systematically validate these models by generating a significant number of both positive and negative examples.[44] Continued efforts to carefully collect and curate synthesis information from the literature will also be paramount to the development of improved models for synthesis.

## Methods

### Restructuring text-mined data

We examined a subset of the text-mined dataset of solid-state inorganic reactions constructed by Kononova *et al*. to calibrate $\Gamma$ on successful reactions.[27] The dataset itself required some restructuring and additional computation to this end. We filtered for reactions targeting ternary oxides in MP taking place in an air environment. To simplify the calculation of $\Gamma_{opt}$, we considered only reactions having no more than two solid precursors (with all precursors listed in MP) and only $O_2$, $CO_2$, and $H_2O$ as allowed byproducts or gaseous precursors. If no environment was given, the environment was assumed to be air. The synthesis temperature was identified as the maximum temperature used in a heating or sintering step. If no temperature was given, the temperature was assumed to be 1073 K.

We used pre-calculated DFT energies (GGA/GGA+U)[45] from MP for the enthalpies of formation and $E_{hull}$, assuming that the composition listed as a precursor or target in the text-mined dataset was in the lowest energy structure for that composition in MP. We used the model proposed by Bartel *et al*. to determine the Gibbs energy of solid phases at the listed reaction temperatures.[41] NIST data was used for the participating gasses, $O_2$, $CO_2$, and $H_2O$.[46]

**Calculating thermodynamic selectivity**

We implemented a modified version of the *reaction-network* package,[25,26] available at https://github.com/materialsproject/reaction-network. Our own reaction generation workflow takes in as arguments target material and temperature and requires three steps. First, a reaction network is generated at 300 K for all materials in the target's chemical space that are in MP and labeled as experimentally observed, specifying the set of allowed precursors as all precursors used in the filtered text-mined reactions. The thermodynamics associated with reactions in the network are then updated based on the synthesis temperature of interest. Finally, target-forming reactions in the network are identified as all reactions that lead to the desired target without solid byproducts. The resulting reaction networks were used to compute $\Delta G_{rxn}$, $C_1$, and $C_2$ as described in Ref. [25]. These quantities were combined into the aggregate quantity $\Gamma$ according to **Equation 1**.

The weights ($w_0$, $w_1$, $w_2$) in this expression were chosen arbitrarily in Ref. [25]. Under the assumption that the optimum (lowest) $\Gamma$ should correspond with the most likely reaction, we re-optimized the weights $w_0$, $w_1$, and $w_2$ by minimizing the $L_2$ loss between the lowest $\Gamma$ ($\Gamma_{opt}$) and the $\Gamma$ associated with the observed reaction ($\Gamma_{obs}$). The sequential least-squares programming (SLSQP) algorithm, as implemented in scipy,[47] was applied to this end, with multiple initializations. Constraints requiring that the three weights sum to 1 and that each lies within [0, 0.99] were imposed, as prescribed by Ref. [25]. The set of weights yielding the lowest objective value was selected, resulting in the updated functional form of $\Gamma$ given by **Equation 2**.

$$\Gamma = 0.16\,\Delta G_{\mathrm{rxn}} + 0.56\,C_1 + 0.28\,C_2 \tag{2}$$

As this reweighting approach remains somewhat arbitrary, other sets of weights were considered to ensure that the proposed bounds are not highly sensitive to the weighting choice. In **Figure S4**, the cumulative distributions of $\Gamma_{obs}$ and $\Gamma_{opt}$ are shown for three cases: (a) the weights proposed by McDermott *et al.* ($w_0$ = 0.1, $w_1$ = 0.45, $w_2$ = 0.45), (b) those associated with equal weighting ($w_0 = w_1 = w_2 = 0.33$), and (c) those used in this work ($w_0$ = 0.16, $w_1$ = 0.56, $w_2$ = 0.28). Regardless of weight choice, 90% or more of the text-mined reactions exhibit $\Gamma < 0.1$ eV/atom.

For novel materials with unknown synthesis recipes, we generated reactions at 600, 900, 1200, 1500, and 1800 K and chose the "optimum" reaction as that with the minimum $\Gamma$ across all temperatures.

We sought to further corroborate two approximations in this workflow on a small subset of the generated materials: (1) the consideration of only text-mined precursors in reaction enumeration and (2) the consideration of only the aforementioned temperatures. We validated (1) by assessing $\Gamma_{opt}$ with an expanded precursor set for the 44 generated targets in the Ba-Cu-O chemical space. In addition to text-mined precursors, we allowed for experimentally realized materials listed in the Materials Project with $E_{hull} < 50$ meV/atom. This expanded our allowed precursor set from 14 to 18 materials while still ensuring that the chosen set of precursors were plausibly accessible materials rather than materials where a new synthesis recipe would need to be developed to obtain them. In **Figure S5a**, we compare the $\Gamma_{opt}$ associated with the expanded precursor set vs. that associated with the text-mined precursor set. The addition of new precursors does not change $\Gamma_{opt}$ for 23 of 44 targets. For the remaining targets, $\Gamma_{opt}$ changes negligibly for most cases, and importantly never decreases by more than the proposed 100 meV/atom bound. We validated (2) by assessing $\Gamma_{opt}$ for the same 44 targets in the Ba-Cu-O chemical space using a finer discretization of 100 K. In **Figure S5b**, we show a parity plot of $\Gamma_{opt}$ using each of the two temperature intervals and find that discretizing every 300 K gives nearly identical results.

**Novel materials generation**

Novel materials were generated using the generative model Chemeleon as detailed at https://github.com/hspark1212/chemeleon.[36] We considered novel materials in the same chemical spaces listed in the text-mined dataset so that we could apply the known text-mined precursors for reaction generation. Hundreds of formulas were enumerated for each chemical space, and 50 formulas per chemical space were randomly selected from this larger pool of compositions. Chemeleon takes a formula string as input and generates structures accordingly. We relaxed the output structures using pre-trained CHGNet[11] version 0.3.0 to a force cutoff of 0.1 eV/Å. Relaxations were performed using the Atomic Simulation Environment.[48] The ground-state structure associated with each formula was identified accordingly and used for the calculation of $E_{hull}$ and $\Gamma_{opt}$. The chemical spaces with the greatest number of materials having $E_{hull} < 100$ meV/atom were considered for this examination, resulting in 2673 generated materials spanning 88 ternary oxide chemical systems. Note that while we used CHGNet for novel material energetics, the text-mined material energies were drawn from GGA(+U) calculations in MP. Prior work has shown that CHGNet may slightly underestimate $E_{hull}$ when applied in this manner for generated materials.[49] We employed pymatgen to generate phase diagrams for stability determination.[50]

For validation of thermodynamic quantities derived from CHGNet, 100 generated materials were randomly selected for relaxation with DFT using MP-compatible settings at the GGA(+U) level of theory.[41,47] The 0 K formation energies and temperature-dependent reaction energies were subsequently recomputed. In **Figure S6**, we show the performance of CHGNet relative to DFT for these 100 materials, comparing formation energies, $E_f$ (a) and reaction energies $\Delta G_{rxn}(T)$ (b) computed with CHGNet total energies for the target material vs. with DFT total energies for the target material. The corresponding distributions of errors are shown in **Figure S6c-d** along with the mean absolute error (MAE) and mean error (ME). The energies from the two sources are highly correlated ($R^2 \geq 0.98$), but CHGNet underpredicts relative to DFT, with mean errors of ~25-35 meV/atom. Errors of this magnitude do not alter the overall trends reported in this work. Further, the more negative energies predicted by CHGNet suggest that the reported $E_{hull}$ and $\Gamma_{opt}$ would be slightly larger if recomputed with DFT, exacerbating the discrepancies between most model predictions and these thermodynamic heuristics.

**Data-driven synthesizability predictors**

For all four ML synthesizability prediction models, we used the pre-trained models provided with each publication. New materials were evaluated via a forward pass through the trained models. Consistent with these publications, we considered a classification threshold score of 0.5 for each model, wherein materials assigned scores $> 0.5$ are classified as synthesizable and those assigned scores $< 0.5$ are classified as unsynthesizable.

PU-CGNF[30], a compositional model, was trained and validated on previously synthesized (positively labeled) and as-yet-unsynthesized (unlabeled) materials from MP and OQMD. Unlabeled materials were assigned hard labels *via* a transductive bagging support vector machine (SVM). With this method, in each iteration a different subset of the unlabeled materials is assigned "negative" labels to train an SVM together with the positive-labeled materials, and the SVM is subsequently applied to the rest of the unlabeled materials. SVM predictions from all iterations are aggregated to assign final labels to the entire set of unlabeled materials. A formula string given to PU-CGNF for classification is encoded with *Roost*[51] and passed through the neural networks associated with each bagging iteration. The scores output from each classifier are averaged, yielding a final score corresponding to the likelihood of falling into the synthesizable class. The implementation for PU-CGNF is further detailed at https://github.com/kaist-amsg/Synthesizability-stoi-CGNF/.

SynthNN,[31] also compositional, was trained on (positive-label) experimentally synthesized materials from the ICSD[16] and unlabeled data generated artificially. During training, unlabeled formulas are first treated as negative examples such that the likelihood of each formula falling into the positive or negative class can be determined. Then, unlabeled materials are duplicated such that each is labeled as both positive and negative. The formula associated with each classification (positive or negative) is weighted during subsequent training by the newly determined likelihood of it falling into that class, such that the higher-likelihood label has a greater influence in training. A formula string given to SynthNN for classification is encoded using the Atom2Vec compositional representation,[52] wherein composition is represented as a vector of the relative atomic fraction associated with each element type. This material encoding is passed through a deep neural network trained on the positive and newly labeled materials, which yields a score corresponding to the likelihood of falling into the synthesizable class. The implementation for SynthNN is further detailed at https://github.com/antoniuk1/SynthNN/.

PU-CGCNN[29] employed the same PU-learning approach as PU-CGNF during training, but takes crystal structure as input, rather than composition alone. PU-CGCNN uses a crystal graph convolutional neural network (CGCNN) architecture[53] for material encoding and classification. In addition, PU-CGCNN is trained on data from MP but not OQMD, resulting in ~300,000 fewer unlabeled materials compared with PU-CGNF. A material, as defined by its composition and structure, given to PU-CGCNN for classification is passed through the CGCNN associated with each bagging iteration. The scores output from each classifier are averaged, yielding a final score corresponding to the likelihood of falling into the synthesizable class. The implementation for PU-CGCNN is further detailed at https://github.com/kaist-amsg/Synthesizability-PU-CGCNN/.

SynCoTrain[32] is another structure-based model trained on oxide materials from MP. SynCoTrain employs a co-training strategy with two distinct end-to-end material representation and property prediction architectures, ALIGNN[54] and SchNetPack,[55] alternating between architectures in each iteration. In each iteration, a bagging SVM is applied to assign labels. Then, newly assigned positive labels are assimilated into the existing positive dataset for subsequent training steps. Ultimately, remaining labels are assigned from the model iteration with the best recall. A material, as defined by its composition and structure, given to SynCoTrain for classification is encoded and classified using a new SchNet model trained on the positive and newly labeled materials, which yields a score corresponding to the likelihood of falling into the synthesizable class. The implementation for SynCoTrain is further detailed at https://github.com/BAMeScience/SynCoTrainMP.

The predictions made by all models on the Chemeleon-generated materials are provided in the **Supplementary Information** along with their computed $E_{\text{hull}}$ and $\Gamma_{\text{opt}}$.

**Data and Code Availability**

The workflow applied in this work, as well as the data used in our analysis, are available at the following GitHub repository: https://github.com/Bartel-Group/synth_assess.

## Acknowledgements

JS and CJB acknowledge support from National Science Foundation Award No. 2435911 and the University of Minnesota College of Science and Engineering Data Science Initiative. SH acknowledges support from the University of Minnesota Undergraduate Research Opportunities Program (UROP). The authors acknowledge the Minnesota Supercomputing Institute (MSI) at the University of Minnesota for providing resources that contributed to the research results reported herein.

**Supplementary Information**

**Thermodynamic assessment of machine learning models for solid-state synthesis prediction**

Jane Schlesinger[1], Simon Hjaltason[1], Nathan J. Szymanski[1], Christopher J. Bartel[1*]

[1] University of Minnesota, Chemical Engineering and Materials Science, Minneapolis, MN 55455

[*] correspondence to cbartel@umn.edu

**Table S1.** Target formula, reaction string, temperature (K), $E_{hull}$ (eV/atom), $\Delta G_{rxn}$ (eV/atom), $C_1$ (eV/atom), $C_2$ (eV/atom), $\Gamma_{obs}$ (for the listed reaction and temperature, eV/atom), $\Gamma_{opt}$ (for the listed target and temperature, eV/atom) for each text-mined reaction considered here. The full dataset is provided in the accompanying file table_s1.csv.

**Table S2.** Target formula, $E_{hull}$ (eV/atom), $\Gamma_{opt}$ (eV/atom), and synthesizability predictions for each of the ML predictors for each generated material considered here. $E_{hull}$ and $\Gamma_{opt}$ were computed using CHGNet-calculated energies for the generated material and DFT-calculated materials from the Materials Project for the other materials in the chemical space. The full dataset is provided in the accompanying file table_s2.csv.

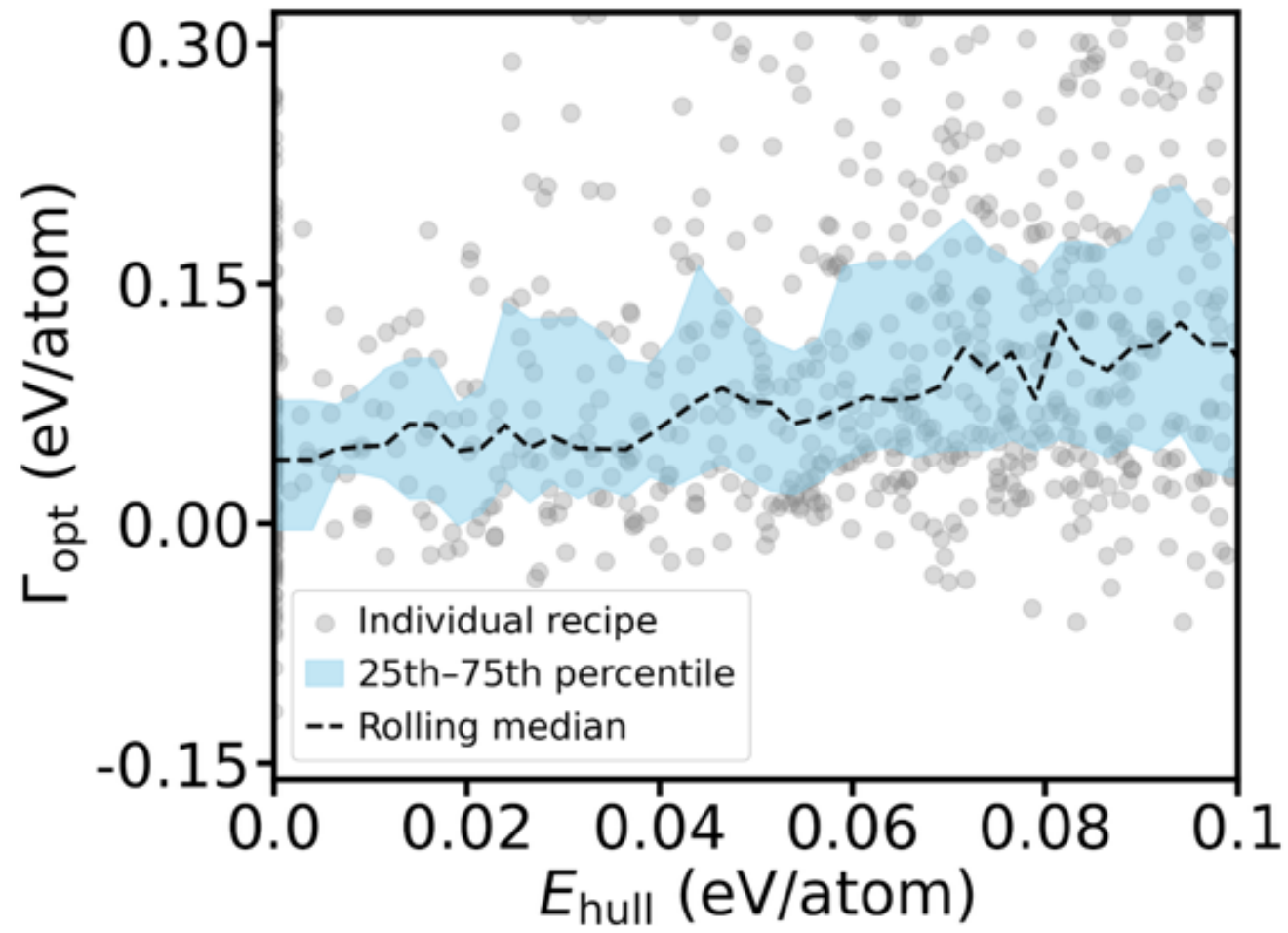

**Figure S1.** $\Gamma_{opt}$ vs. target $E_{hull}$ for each Chemeleon-generated materials (grey points) and a rolling median of $\Gamma_{opt}$ as a function of target $E_{hull}$, with the median evaluated over a range of 8 meV/atom every 2.5 meV/atom. The distance from the median to the 25th and 75th percentile values of $\Gamma_{opt}$ is shaded in light blue. Material energetics were computed using CHGNet,[11] discussed further in **Methods**.

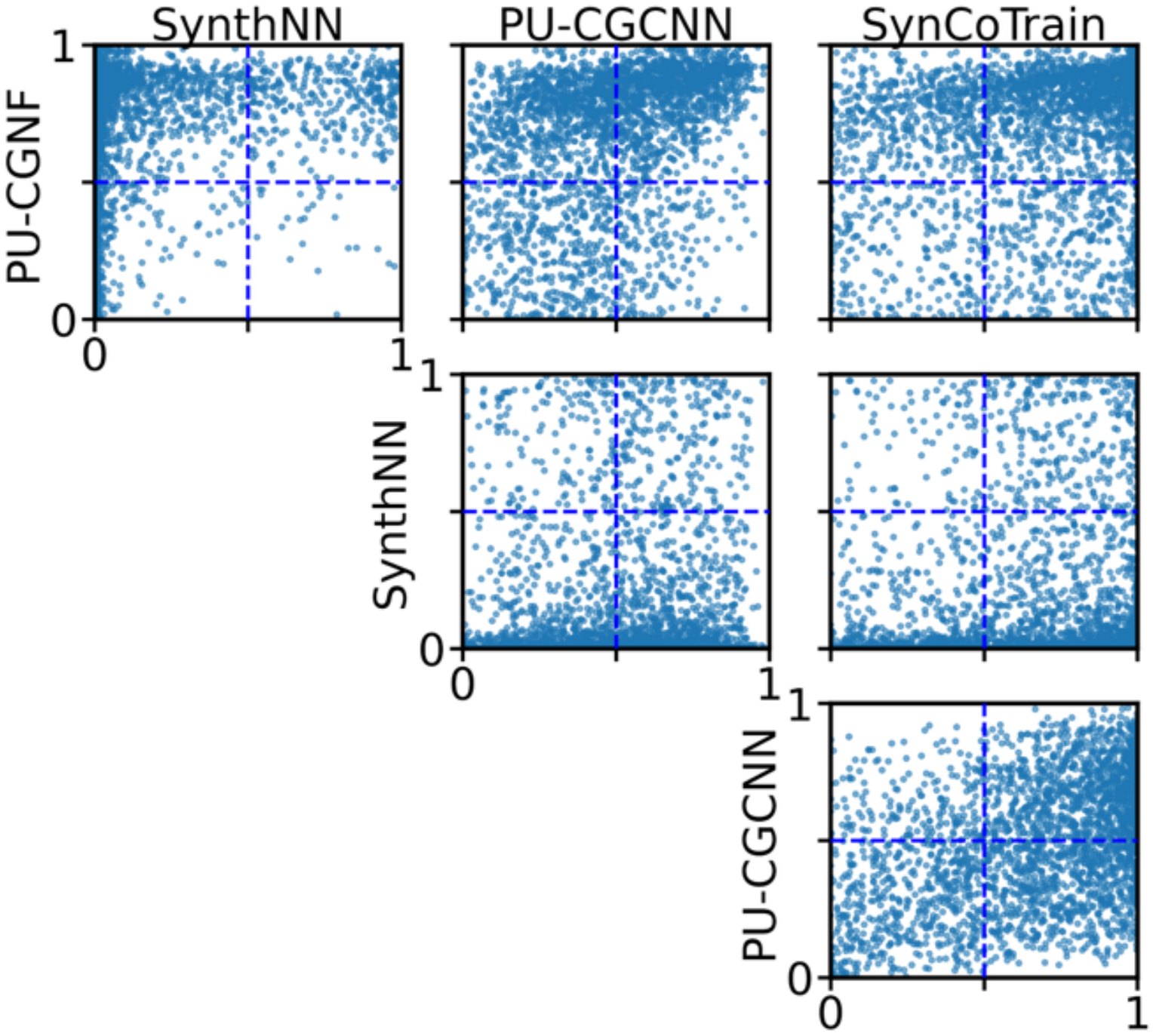

**Figure S2.** Pairwise score comparisons for each of the four ML predictors compared to one another. Each point represents a generated material, with its coordinates representing the score ascribed to it by each model. The blue dashed lines indicate the score cutoffs for synthesizability classification (0.5), such that points that fall in the lower-left and upper-right quadrants are materials classified as unsynthesizable and synthesizable, respectively, by both models.

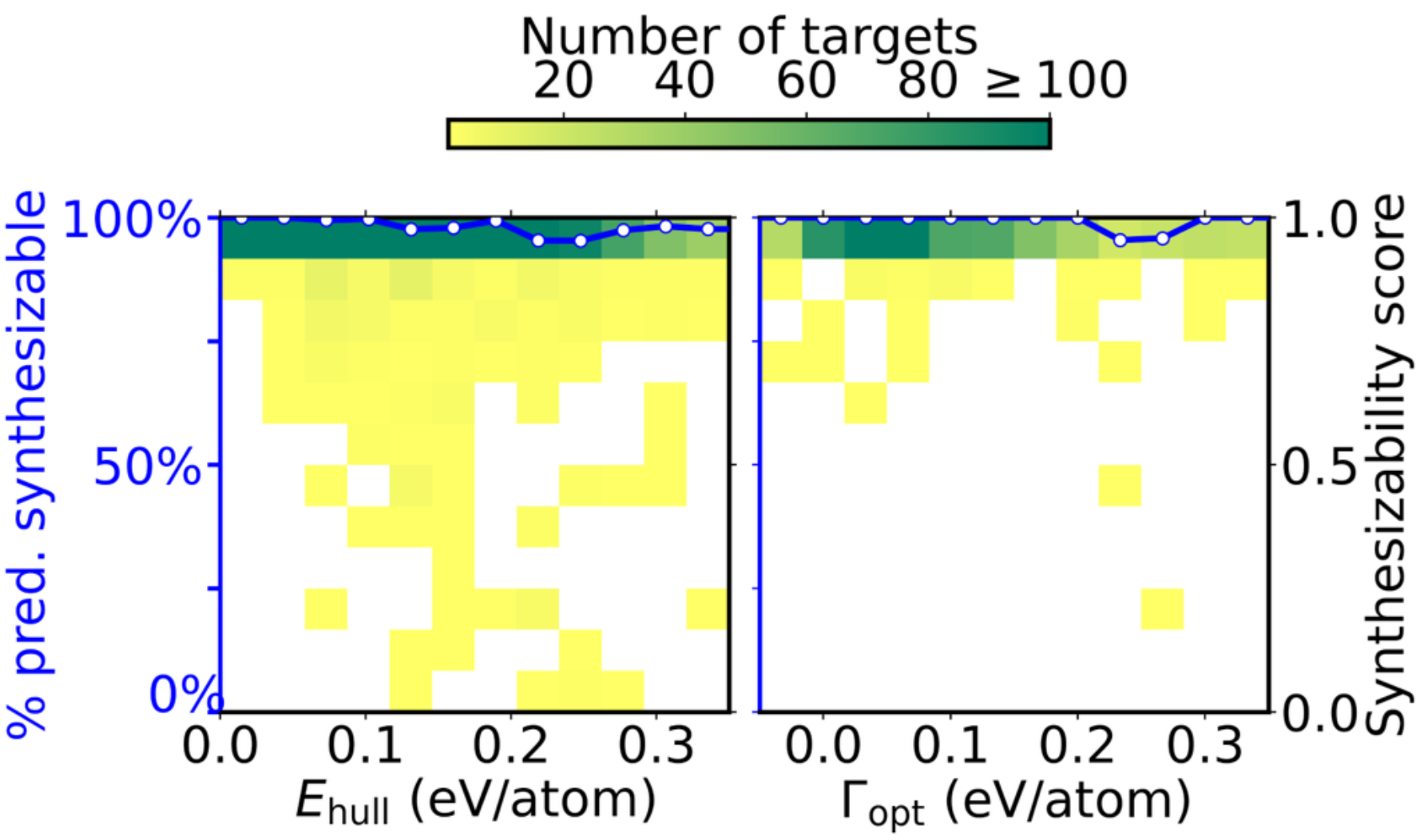

**Figure S3.** TSDNN predictions and CHGNet-calculated $E_{hull}$ (left) and $\Gamma_{opt}$ (right) for Chemelon-generated materials. $\Gamma_{opt}$ is shown only for the 1057 materials with $E_{hull} < 100$ meV/atom. The heatmap color indicates the number of these target materials having a certain synthesizability score (right-hand *y*-axis) and $E_{hull}$ or $\Gamma_{opt}$ (*x*-axis). Moving averages of the fraction of materials predicted synthesizable (score > 0.5) as a function of $E_{hull}$ and $\Gamma_{opt}$ are shown by the blue lines and corresponding left-hand *y*-axes. The moving averages were computed over 30 (left) and 33 (right) meV/atom intervals. White squares indicate no targets falling in a given region.

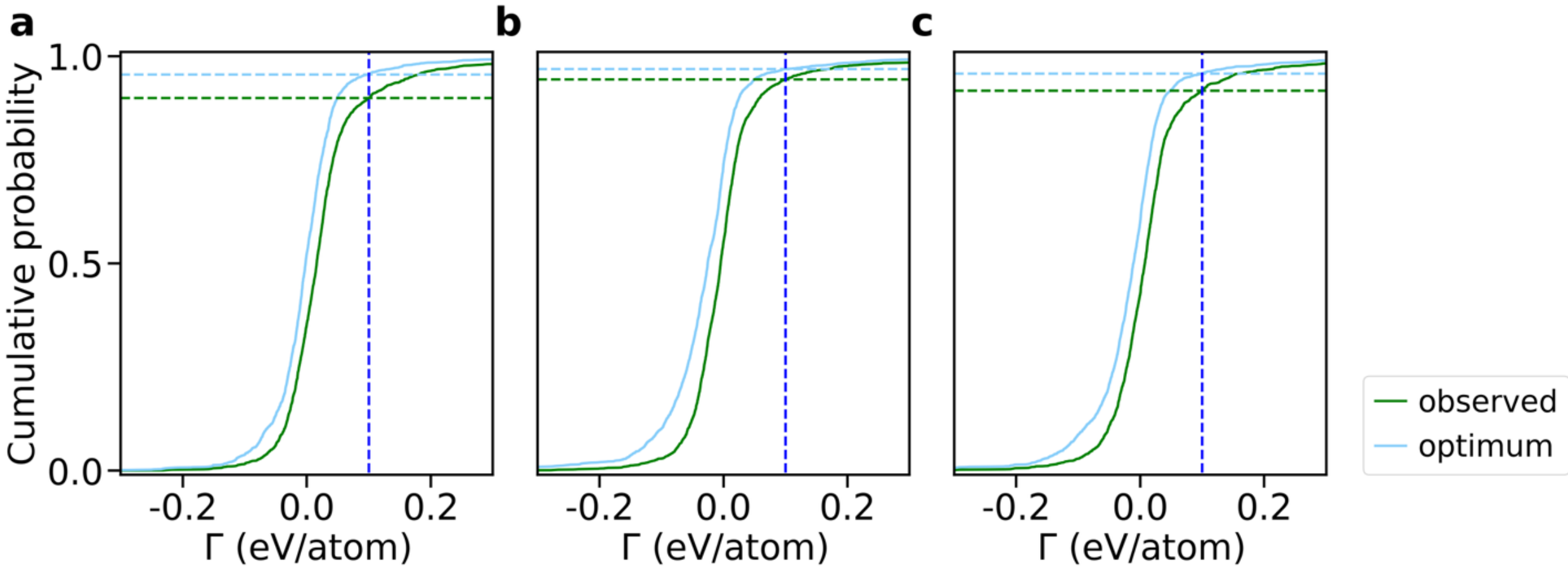

**Figure S4:** Cumulative distributions of Γ computed with **(a)** McDermott *et al.*'s proposed weights ($w_0$ = 0.1, $w_1$ = 0.45, $w_2$ = 0.45). 90% of observed reactions have Γ < 0.1 eV/atom and 96% of optimized reactions have Γ < 0.1 eV/atom. **(b)** Equal weights on each quantity ($w_0 = w_1 = w_2$ = 0.33). 94% of observed reactions have Γ < 0.1 eV/atom and 97% of optimized reactions have Γ < 0.1 eV/atom. **(c)** Our "optimized" weights ($w_0$ = 0.16, $w_1$ = 0.56, $w_2$ = 0.28). 92% of observed reactions have Γ < 0.1 eV/atom and 96% of optimized reactions have Γ < 0.1 eV/atom. In each panel, the vertical blue dashed line indicates Γ = 100 meV/atom, the dashed green horizontal line indicates the cumulative probability of $\Gamma_{obs}$ < 100 meV/atom, and the dashed light blue horizontal line indicates the cumulative probability of $\Gamma_{opt}$ < 100 meV/atom.

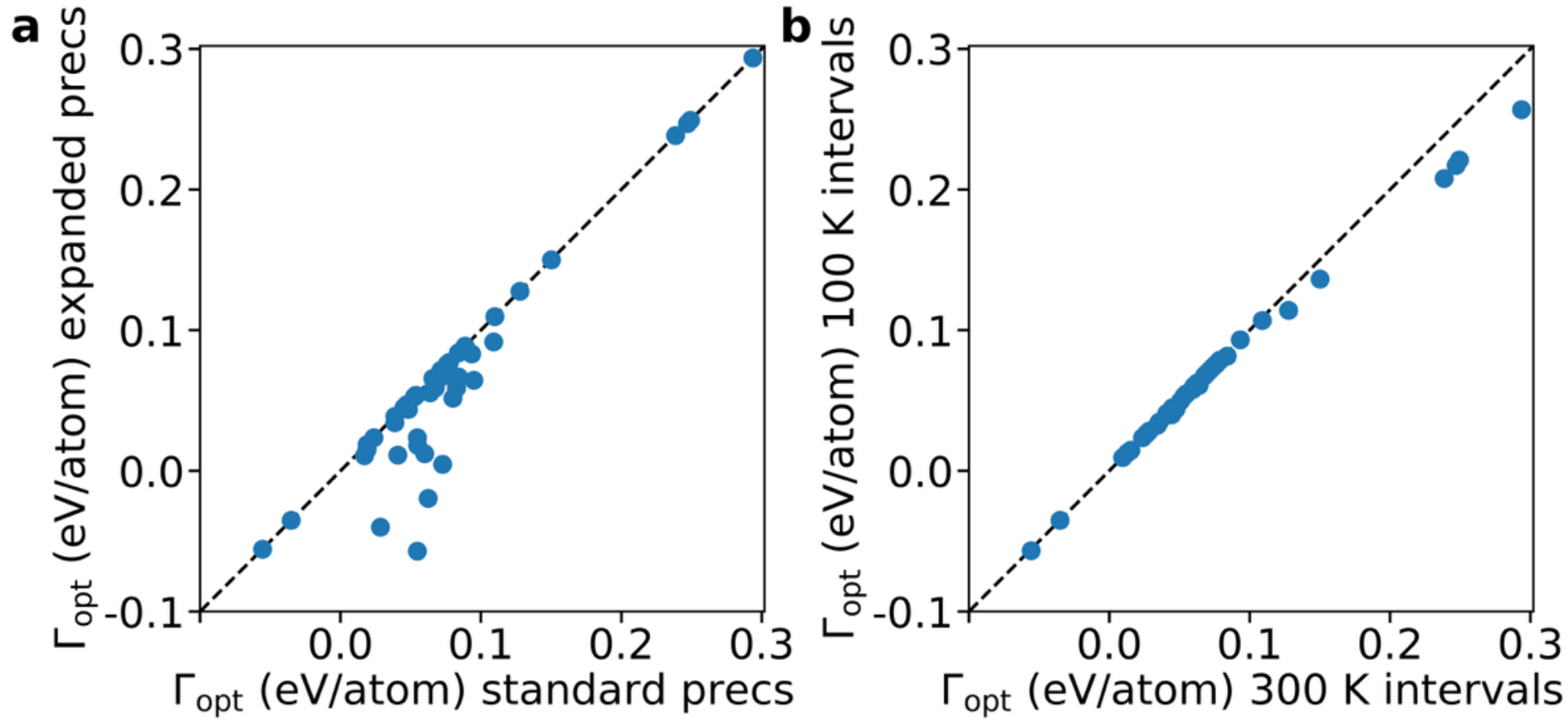

**Figure S5:** Parity plots of the minimum (optimum) Γ for each of the 44 generated targets in the Ba-Cu-O chemical space, considering **(a)** reactions with only precursors for the text-mined dataset as allowed precursors (14 precursors) vs. those with allowed precursors including experimentally realized materials listed in MP within 50 meV/atom of the hull (18 precursors) and **(b)** reactions over 100 K temperature intervals between 300 and 1800 K vs. reactions over 300 K intervals between 600 and 1800 K.

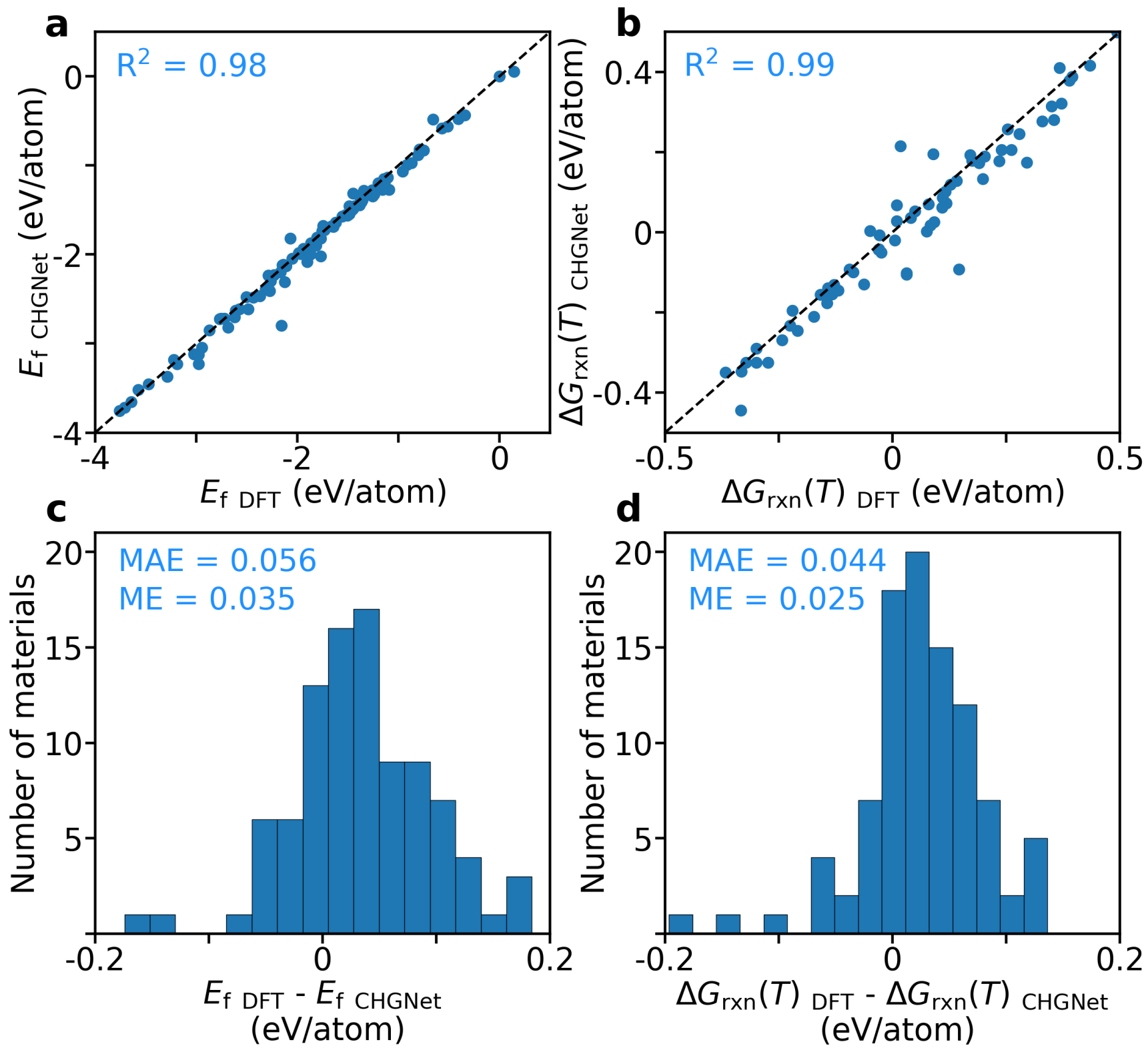

**Figure S6:** A comparison of heuristics determined using CHGNet vs. using DFT for 100 randomly selected materials from the dataset of generated materials. **(a)** A parity plot of $E_f$ computed using a CHGNet total energy for the target material vs. using DFT with GGA/GGA+U functionals for each material with the $R^2$ associated with the plot listed. **(b)** A parity plot of $\Delta G_{rxn}(T)$ for the minimum-Γ reaction computed using a CHGNet total energy for the target material vs. using DFT with GGA/GGA+U functionals for all relevant species with the $R^2$ associated with the plot listed. **(c)** The distribution of error $E_{f\,DFT}$ - $E_{f\,CHGNet}$ for each material and associated mean absolute error (MAE) and mean error (ME) for the whole dataset listed. **(d)** The distribution of error $\Delta G_{rxn}(T)_{DFT}$ - $\Delta G_{rxn}(T)_{CHGNet}$ for each material and associated MAE and ME for the whole dataset listed.